\documentclass[%
 aip,
rsi,
 amsmath,amssymb,
 reprint,%
]{revtex4-1}

\usepackage{graphicx}
\usepackage{dcolumn}
\usepackage{bm}
\usepackage{xcolor} 
\usepackage[utf8]{inputenc}
\usepackage[T1]{fontenc}
\usepackage{mathptmx}
\usepackage{etoolbox}
\usepackage{ulem}

\makeatletter
\def\@email#1#2{%
 \endgroup
 \patchcmd{\titleblock@produce}
  {\frontmatter@RRAPformat}
  {\frontmatter@RRAPformat{\produce@RRAP{*#1\href{mailto:#2}{#2}}}\frontmatter@RRAPformat}
  {}{}
}%
\makeatother
\begin{document}

\preprint{AIP/123-QED}
\title{A High Resolution Dilatometer Using Optical Fiber Interferometer}

\author{Xin Qin} 
\affiliation{International Center for Quantum Materials,
  Peking University, Haidian, Beijing 100871, China}
\affiliation{Hefei National Laboratory, Hefei 230088, China}

\author{Guoxin Cao}
\affiliation{College of Underwater Acoustic Engineering, Harbin Engineering University, Harbin 150001,China}%
\affiliation{Acoustic Science and Technology Laboratory, Harbin Engineering University, Harbin 150001,China}

\author{Mengqiao Geng} 
\affiliation{International Center for Quantum Materials,
  Peking University, Haidian, Beijing 100871, China}
\affiliation{Hefei National Laboratory, Hefei 230088, China}

\author{Shengchun Liu}
\affiliation{College of Underwater Acoustic Engineering, Harbin Engineering University, Harbin 150001,China}%
\affiliation{Acoustic Science and Technology Laboratory, Harbin Engineering University, Harbin 150001,China}

\author{Yang Liu}
\email{liuyang02@pku.edu.cn}
\affiliation{International Center for Quantum Materials,
  Peking University, Haidian, Beijing 100871, China}
\affiliation{Hefei National Laboratory, Hefei 230088, China}

\date{\today}

\begin{abstract}
 We introduce a high performance differential dilatometer based on an all-fiber Michelson interferometer at cryogenic temperature with $10^{-10}$ resolution in $\delta L/L$. It resolve the linear thermal expansion coefficient by measuring the oscillating changes of sample thickness and sample temperature with the interferometer and in-situ thermometer, respectively. By measuring the linear thermal expansion coefficient $\alpha$ near the antiferromagnetic transition region of BaFe$_2$As$_2$ as a demonstration, we show our dilatometer is able to measure thin samples with sub-pm-level length change resolution and mK-level temperature resolution. Despite there is residual background thermal expansion of a few nm/K in measurement result, our new dilatometer is still a powerful tool for study of phase transition in condensed matter physics, especially has significant advantages in fragile materials with sub-100$\mu$m thickness and being integrated with multiple synchronous measurements and tuning thanks to the extremely high resolution and contactless nature. The prototype design of this setup can be further improved in many aspects for specific applications.
\end{abstract}

\maketitle

\section{\label{introduction}INTRODUCTION}

Thermal expansion coefficient $\alpha$, the thermodynamical parameter
that describes the relation between sample size (or volume) and
temperature, carries crucial information about materials. Studying
$\alpha$ across phase transition is of particular interest in condensed
matter physics, since it is coupled to all phase transitions allowed
by symmetry and can be linked to other parameters such as the
specific heat $C_\mathrm{V}$ through the Ehrenfest and Maxwell
relations\cite{RN202}. The thermal expansion measurement
can detect phase transitions and dynamics properties of superconductors\cite{RN226,RN227,RN228,RN229,RN230}, magnetic materials\cite{RN231,RN232,RN233,RN234} and even metallic glass and so on, which shows its unique advantages
compared to resistivity and magnetic-susceptibility measurements.

Dilatometers measure the sample length $L(T)$ as a function of
temperature $T$, the variation $\Delta L = L(T)-L(T_0)$ from its value
at a reference temperature $T=T_0$, and then deduce the thermal
expansion coefficient $\alpha=L^{-1} dL/dT$. Different approaches of
measuring $\Delta L$ have been reported, using fiber Bragg grating\cite{RN223,RN224}, atomic microscope piezo-cantilever \cite{RN206}, piezobender \cite{RN207}, strain gauge
technique \cite{RN208,RN225}, X-ray diffractions \cite{RN209}, etc. The
capacitive dilatometry, which measures $\Delta L$ through monitoring
the capacitance between the sample surface and a metal reference
surface, is one of the most widely used because of its supreme
accuracy\cite{RN210,RN211,RN212,RN213,RN222}. Its pm-resolution
of $\Delta L$ corresponds to $\Delta L/L\sim 10^{-10}$ for samples
with a length $L \leq 5$ mm, orders of magnitudes better than other
methods.

Here we present a new approach which measures the samples' length
variation contactlessly with a sensitivity of a few $10^{-10}$ in $\delta L/L$ for sub-milimeter-thickness sample using a high resolution optical fiber Michelson interferometer. We use a "true" differential method where the applied quasi-DC sinusoidal temperature oscillation $\delta T$ at $f_S$ near $T_0$ induces an oscillation $\delta L$ in $L$. So $\delta L$ naturally has in-phase and out-of-phase components refer to $\delta T$, which are from the thermal transport delay because of sample thermal propeties. When the sample is at equilibrium
and has a uniform phase under AC temperature oscillation $\delta T$, there is no phase shift between $\delta L$ and $\delta T$ and the out-of-phase component is zero. The amplitudes of $\delta L$ and $\delta T$, $\langle \delta L\rangle$ and $\langle \delta T \rangle$, can be measured using lock-in technique to deduce the linear thermal expansion coefficient $\alpha=L^{-1} \langle \delta L\rangle / \langle \delta T\rangle$. This setup can achieve extraordinary 0.5 pm resolution in
$\langle \delta L\rangle$, $5 \times 10^{-10}$ in $\delta L/L$ for 1-milimeter-thickness sample, and mK-level resolution in $\langle \delta T\rangle$, simultaneously.  
 Thanks to the contactless
nature of our optical approach, we can study fragile and thin samples,
implement other synchronous measurements such as transport, as well as
apply in-situ controls such as strain and electric field. In this work we demonstrate our dilatometer at liquid nitrogen temperatures while it can, in principle, be extended to temperatures ranging from sub-10 K up to room temperature. 

This article is organized in the following order. We first describe
the principle and design of our dilatometer and the extension for ultra-low temperature in Section II. We then
calibrate its resolution, stability and precision in Section III. In
Section IV, we apply this setup to study the phase transition of a thin BaFe$_2$As$_2$
crystal as a demonstration.

\section{\label{PRINCIPLE AND SETUP} Setup and Principle}

\subsection{\label{Dilatometer Structure}Dilatometer Structure}

Fig. 1 shows the structure of our dilatometer. In general, we design a setup which contains a fiber Michelson interferometer to measure $\delta L$ and in-situ heater and thermometer to apply and measure $\delta T$ of sample. As shown in Fig. 1(a), the device is attached
to the cold finger inside the cryostats through the upper screw thread
(label 1). The beam splitter of the Michelson interferometer,
i.e. the 50/50 fiber coupler (see next section), is installed inside
the notch of fiber coupler holder (3), and its two 1-meter-length interference arms are winded on
the two PZT rings (5). The whole structure is stacked on the PZT
holders (4) and tightly fixed by the lock nut (2). The diming frame
(7) is mounted to the cold finger (1), inside which the grin-lens
(11) is installed coaxially using the lens locker (12 and 13), see the
enlarged plot in Fig. 1 (b). The sapphire sample holder (15) is
mounted to an adjustable plate (10) positioned using three Ti fine adjusting screws (6) and springs (8). The coaxial and miniaturized design ensures the temperature uniformity within the structure and minimizes (internal)
vibration. Its 30-mm-diameter and 70-mm-height size fit to most
cryostats.

\begin{figure}[!htbp]
\includegraphics[width=0.45\textwidth]{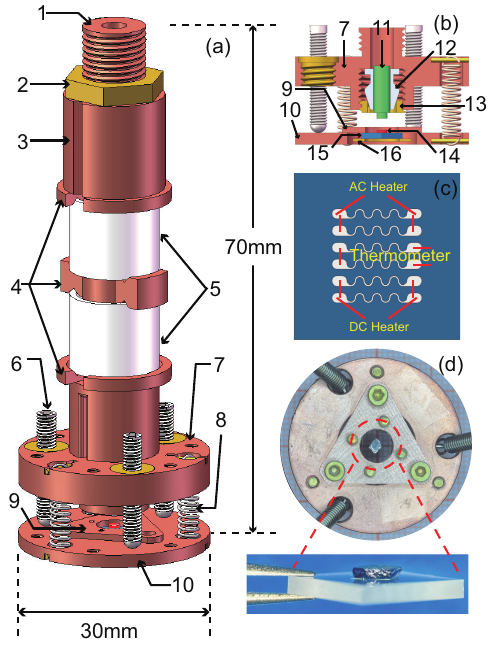}
\caption{\label{Fig1}Schematic drawing and photos of our dilatometer. (a) shows the entire structure, (b) shows a zoom-in cut-away view of the optical adjusting
structure and sample-holding structure, (c) shows the design
of in-situ thermometer, AC and DC heaters on the
sapphire sample holder using six evaporated Pt wires. (d) shows a top-view photo of sample-holding structure with a sample (14) installed on the upper surface of sapphire sample holder (15) by extremely thin N-Grease, about 5 $\mu$m as measured, as shown in the zoom-in photo of sample installation. (1) cold finger, (2)
lock nut, (3) fiber coupler holder, (4) PZT rings holders, (5) PZT
rings, (6) Ti fine adjusting screws, (7) Dimming frame, (8) Be-Cu
springs, (9) upper plate, (10) adjusted plate, (11) grin-lens, (12)
and (13) lens locker, (14) sample, (15) sapphire sample holder, (16) Be-Cu
lower plate. The structure component is made of oxygen-free-copper if not
specifically stated.}
\end{figure}
 
We carefully arrange the upper and lower plates (9 \& 15) and the
sapphire sample holder (15), so that the lower surface of the sample
(14) is perfectly aligned with the upper-surface of the adjustable
plate, which is exactly the plane defined by the pivot of
the three Ti fine adjusting screws (6). With this coplane design, we get rid of the background length change component from thermal expansion of sample-holding structure (9), (10), (15), (16) so that the background AC length change at measurement frequency $f_S$ of entire system, $\delta L_B$, is limited to the distance change along the Z direction between the end of grin lens (11) and the pivot of three Ti fine adjusting screws (6), which is from the AC thermal expansion of 10-mm-long Ti fine adjusting screws (6), grin lens (11) and lens-holding structure (7), (12), (13).

Then we thermally isolate structures from sample by introducing $\delta T$ only at sample and making sure it doesn't leak to fine adjusting screws (6) as well as other structures for further surpressing this background AC length change so that the length change measured by the interferometer origins only from the sample's thermal expansion.
Firstly, we cool the entire structure to a low temperature and
stabilize it with a system PID temperature controller using the cryostat
thermometer and heater mounted on the cold finger. Then we introduce $\delta T$ at sample directly by installing sample on the sapphire sample holder (15), which has six Pt wires evaporated on its lower surface as in-situ heaters and thermometer, as shown in Fig. 1(c). Two pairs of Pt wires are used as DC and AC heaters that the voltage $V_{DCH}$ applied to the DC heater heats the
sample temperature to $T_0$, and an oscillating voltage $V_{ACH}$ at $f_S$
applied to the AC heater induces an oscillation $\delta T$ to the
sample DC temperature $T_0$. $\delta T$ leaks to Ti fine adjusting screws (6) through (9), (10) and three contact points between (6) and (10) and the thermal resistance and thermal capacity are large enough to reduce the $\delta T$ at (7) is zero at a normally used AC heating frequency $f_S$, i.e. $f_S>0.01$Hz, so the background AC length change at $f_S$ is limited to the thermal expansion of 10-mm-long Ti fine adjusting screws (6) and our in-situ AC heating can also reduce the $\delta T$ heating of fine adjusting screws (6) comparing with AC heating the entire structure. The central pair of Pt wires are calibrated by the cryostat thermometer and used to monitor the real-time sample
temperature $T_0+\delta T$ via lock-in technique. What's more, we introduce thermal insulation between the sample holder (15) and the upper and lower plate (9 \& 16), as well as between the fine adjusting screws (6) and the adjusted plate (10). Besides, the adjusted plate (10) is relatively large and thermally shunted to the thermostatical cold finger (1) and the thermal expansion coefficient of Ti is very small. So the AC
length fluctuation of the fine adjusting screws (6) is well reduced with the designs mentioned above.

We can model the temperature variation of the fine adjust screws (6)
and minimize it further by carefully choosing the material/size of each part
and changing the thermal link between different parts. Eventually, the
thermal expansion of the fine adjust screws (6) becomes negligible for
$f_S\gtrsim 0.1$ Hz even if the sample thickness is less than 100
$\mu$m. The thermal equilibrium time of the sapphire sample holder (15)
with thickness $h$ is proportional to $h^2$, and deformation caused by thermal gradient is inversely proportional to $h^3$. Our
numerical modeling shows that if $h=800$ $\mu$m, the temperature
oscillation amplitude at the sapphire plate's upper surface is about
half of the value at its lower surface when the AC heating frequency
is set to about 10 Hz. Giving consideration to other effects such as
the thermal resistance of the sample and between the sample and
sapphire plate, we choose AC temperature oscillation frequency
$f_S \lesssim 2$ Hz to measure $\alpha$.

The in-situ heating is directly applied to the sample holder, so that
the temperature of the rest parts of the dilatometer is stabilized
within mK-level by the system PID temperature controller. Along with
the excellent thermal uniformity among the entire structure and the
minimal mismatch between the interferometer's two fiber arms, the
thermal drift of this interferometer is much smaller and we get better measurement stability comparing with heating the entire structure.

\subsection{\label{Fiber Interferometer} Fiber Interferometer}

As described in the previous section, we coat the upper surface of the sapphire sample holder (15) with an extremely thin layer of N-Grease (typically a few $\mu$m as we measured), and then gently press the sample (14) onto the surface, as the zoom-in photo of sample installation showed in Fig. 1(d). With
the AC heater, we can impose a small temperature oscillation
$\delta T$ at frequency $f_S$ onto a steady state temperature
bias $T_0$ set by the DC heater. The real-time sample temperature
$T_0+\delta T$ can be measured by the thermometer on the lower surface
of sapphire sample holder (15). We then use a fiber Michelson
interferometer to measure the $\delta T$ induced sample's thermal
expansion $\delta L$. Because of the finite specific heat $C$ and thermal conductivity $\kappa$ of the sample and sample holder, there is always a delay $\tau\propto C/\kappa$ between sample temperature and the temperature read from in-situ thermometer. Since the thermal conduction of the thin grease and sapphire sample holder (15) is large, $\tau$ is usually limited by the sample, especially near phase transition region. Therefore, the measured $\delta L$ naturally has in-phase and out-of-phase components refer to measured $\delta T$ and the out-of-phase component has great potential for the simultaneous measurement of specific heat and thermal conductivity of sample around phase transitions\cite{RN235}. Typically the out-of-phase component is non-negligible when $f_S>1$ Hz. If we choose a small $f_S$ so that $f_S\cdot\tau<<1$, we can neglect the out-of-phase component and measure the oscillation amplitude of $\delta T$ and $\delta L$,  $\langle \delta T\rangle$ and $\langle \delta L\rangle$, by digital lock-in technique to deduce the sample's thermal expansion coefficient at $T_0$ by:

\begin{equation}\label{eq1}
\alpha_{T_0}=L^{-1}\frac{\langle \delta L\rangle}{\langle \delta T\rangle}
\end{equation}
Then we sweep $V_{DCH}$ applied to the DC heater to slowly tune DC temperature $T_0$ to sweep the $\alpha$ vs. $T_0$ continuously.

\begin{figure}[!htbp]
  \includegraphics[width=0.45\textwidth]{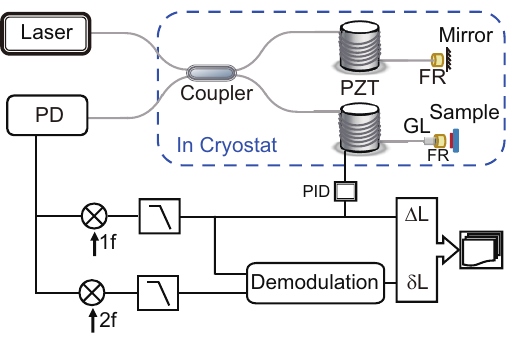}
  \caption{\label{Fig2} Principle of our optical interferometer
    measurement. The AC \& DC change of sample thickness, $\delta L$ \&
    $\Delta L$, is translated to the change of AC and DC component of
    the total phase difference $\Phi$, $\delta \phi_S$ \&
    $\Delta \phi_S$, respectively. These changes can be deduced from the
    photo-detector's (PD's) output voltage. We wind the two arms or the
    fiber Michelson interferometer around two PZT rings tightly in order
    to apply phase modulation and feedback. We deduce the AC phase
    change $\delta L$ from the demodulation output directly, and
    calculate the DC change of sample thickness $\Delta L$ from the
    feedback voltage.}
  \end{figure}

The measurement principle of our all-fiber Michelson interferometer is
shown in Fig. 2. The 10 kHz-linewidth, $\lambda=$1550 nm laser
is equally distributed into the measurement and reference arms by a
2x2 fiber coupler. We use a grin-lens at the end of the measurement
arm to collimate the outgoing laser to be parallel and collect the light reflected from the upper
surface of the sample, while the reference arm has an integrated
mirror at its end. The diameter of laser spot on sample surface is about 450 $\mu$m and the light intensity we use is limited to be smaller than 100$\mu$W. With typical thermal anchoring, such a low light power can be used in cryogenic systems even down to sub-1 K temperatures without noticeable heating. The system signal to noise ratio (SNR) is influenced by the interference contrast of our interferometer, which is dependent on the ratio of collection light intensity of reflected light from sample surface and emition light intensity from grin lens. So for the best resolution we need the sample surface is flat on the scale of our laser spot size for better specular reflection of incidence parallel light
, i.e. the cleavage surface of BaFe$_2$As$_2$ and the polished surface of Ag with a roughness of about 1.5 $\mu$m are enough for almost 100\% specular reflection. Besides we should adjust three Ti fine adjusting screws (6) to make sure the reflected laser returns by the same route of emition laser and completely collected by the grin lens (11). The whole structure is coaxial and with 120°-rotational symmetry so the thermal expansion and vibration of structure don't affect the alignment we adjust initially in the atmosphere when we change the system temperature. The work distance of grin lens is 5 $\sim$ 10 mm so the influence of thermal expansion of structure over measurement temperature range, which is at level of $\mu$m, is negligible. Faraday-Rotation-Lenses (FR) are inserted before the
sample/mirror in order to reduce the random polarization rotation of
the winded fiber\cite{RN200}. The reflected laser beams interfere at
the coupler so that the light intensity $I$ at the output port is
$I=I_0[1+\cos(\Phi)]$, where $I_0$ is the interference intensity and
$\Phi$ is the phase difference between these two fiber arms. The
photodetector (PD) converts $I$ into a voltage signal.

We implement the PGC modulation/demodulation methods to improve the
sensitivity in measuring $\Phi$ \cite{RN201}. We tightly wind
the reference arm around a PZT ring and apply an modulation voltage to
introduce a phase modulation $\phi_M=C\cdot \cos(2\pi f_Mt)$, where
$C$ is the modulation depth and $f_M$ is the modulation frequency
($\sim$ kHz). The modulated light intensity can be expanded according to the Jacobian angular expansion as:
\begin{equation}\label{eq2}
\begin{aligned}
   I&=I_0[1+\cos(C\cdot \cos(2\pi f_Mt)+\Phi)] \\
	&=I_0\left\{ 1+ 2\sin(\Phi)\sum_{N = 0}^{\infty}J_{2N+1}(C)\cos[2\pi\cdot (2N+1) f_Mt]\right. \\
	&\left.+\cos(\Phi)[J_0(C)+2 \sum_{N = 1}^{\infty}(-1)^NJ_{2N}(C)\cos(2\pi\cdot2N f_Mt)] \right\}
\end{aligned}	
\end{equation}
where $J_N$ is the N-order Bessel function. We use lock-in technique
to measure the amplitude $V_{1}$ and $V_{2}$ of the 1$^{st}$ and
2$^{nd}$ harmonic component of $I$ at $f_M$ and $2f_M$, and calculate
$\Phi$ using the relation
\begin{equation}\label{eq3}
\begin{aligned}
  \Phi&=\frac{1}{2}
  \tan^{-1}(\frac{J_2(C)V_{1}}{J_2(C)V_{2}})
\end{aligned}
\end{equation}

$\Phi$ is the quasi-DC component of the total phase difference between
the two arms. It consists a constant value due to the unequal optical
length between the two arms and a oscillating signal from the sample's
thickness oscillation $\delta \phi_S = 4\pi\delta L / \lambda$. It
also has a slowly varying component $\Delta \phi$ caused by the
thermal fluctuation of the fibers and PZTs, wavelength drifting of the
laser, and the thermal expansion of the structure and sample
$\Delta L$ induced by sweeping $T_0$.

In order to compensate the slowly drifting $\Delta \phi$, we
introduce a feedback phase by winding the measurement arm around
another PZT ring and introduce a feedback voltage $V_F$ to it. A
simple feedback voltage can be achieved by filtering and amplifying
the measured $V_{1}$ with an extremely long time constant ($\sim 300 $
s). Once the feedback gain $G_F$, which is now the ratio between the
optical length change and $V_{1}$, is large enough, The DC component
of $V_{1}$ will be approximately zero. Since $\delta \phi_S$ is
usually very small, $V_{1}$ is approximately proportional to
$\delta \phi_S$ and $V_{2}$ is locked at the extremum $I_0J_2(C)$. In
short, by using this feedback, we simplify the measurement, improve
the sensitivity of $\delta \phi_S$, and also solve the diverging
problem caused by the $\tan^{-1}$ function in equation (\ref{eq3}). The
AC length change $\delta L$ can be then simply calculated from
\begin{equation}\label{eq4}
\delta L=\frac{1550}{4\pi J_1(C)I_0}V_{1}	\quad nm
\end{equation}
Meanwhile, the phase induced by the feedback voltage $V_F$ compensates
the quasi-DC change of the sample thickness and the thermal drifts
from laser, fiber and structure. Therefore, if the drifts caused by
the laser, fiber and PZT are small, and the thermal expansion of the
structure $\Delta L_B$ is calibrated by a separate measurement, we can
deduce the absolute change of the sample thickness $\Delta L$ from the
feedback voltage $V_F$ through
\begin{equation}\label{eq5}
\Delta L=G_FV_F-\Delta L_B
\end{equation}

We can also remove sample and evaporate 20-nanometer-thickness gold on the upper surface of sapphire sample holder (15) and it can reflect the incidence laser from grin lens (11) for background thermal expansion measurement. We can get
the structure's DC thermal expansion is $\Delta L_B/\Delta T=G_F V_F/\Delta T$ $\sim$
20 nm/K and its AC thermal expansion
$\delta L_B/\delta T $ is about 4 nm/K at 130K. There is always a background AC thermal expansion $\delta L_B/\delta T$, about a few nm/K, among measurement temperature range and it is slightly different between different measurement cycles because of slightly different sample installation everytime so our prototype dilatometer can't measure the absolute value of $\alpha$ for now if the sample thermal expansion is as small as the background AC thermal expansion $\delta L_B/\delta T$. This background is smooth in temperature, and can be further reduced with better structural design. Fortunately, it does not affect its application in studying phase transitions, where the features in thermal expansion are rather sharp in temperatures. Therefore, our prototype fiber interferometer based dilatometer is a powerful tool in related studies, especially for its extremely high resolution and contactless nature\textcolor{black}{.  }

Before ending this section, we would like to mention that our setup
uses the combination of bend-resistant singlemode fiber and 45 degree
faraday rotators at the end of each arm, in order to minimize the size
\cite{RN200}. Alternatively, one can use polarization-maintaining
fiber for the interferometer when measuring $\alpha$ in a magnetic
field. At a cost, the diameter of the PZT rings should be more than 40
mm.

\subsection{\label{Extension for Ultra-low Temperature} Extension for Ultra-low Temperature}

Although as a prototype we only demonstrate our setup at liquid nitrogen temperature, this method can be extended to ultra-low temperatures. 

The fiber and other optical components still work well in dilution refrigerator below 100 mK from our experience on another work of ultra-low temperature all-fiber MOKE measurement setup so the interferometer should function well. The piezoelectric coefficient of PZT rings decreases with system temperature cooling down and our experience is the piezoelectric coefficient at 1 K is about an order of magnitude smaller than which at room temperature. So feedback is not a problem because we need smaller feedback voltage $V_F$ to cancel the DC drift $\Delta L$ because of the smaller thermal expansion coefficient in ultra-low temperature than room temperature. Besides, we can always identify the modulation depth $C$ from $V_1$, $V_2$ and $I_0$ so the variation of PZT performance doesn't affect the modulation and demodulation mentioned at \ref{Fiber Interferometer}. So the $\delta L$ measurement is extensible for ultra-low temperature. 

As for the $\delta T$ and $T_0$ introducing and measurement, we can still use Pt wires for AC heating and DC heating. Other thermometers such as thermal couple should be used instead of the simple Pt wires. So this setup still works well at ultra-low temperature in principle.

\section{\label{Resolution, Stablization and Precision}Resolution, Stablization and Precision}

\begin{figure}[!htbp]
\includegraphics[width=0.45\textwidth]{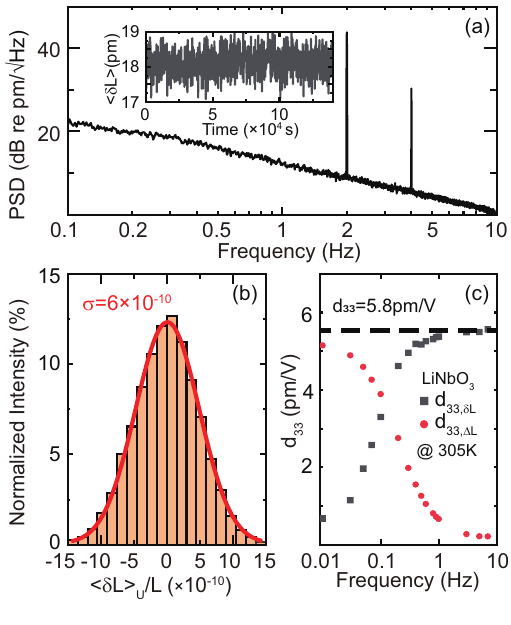}
\caption{\label{Fig3} Calibration of our optical interferometer
  measurement. (a) The Power Spectra Density (PSD) of $\delta L$. An
  AC sample thickness oscillation $\langle\delta L\rangle=18$ pm is induced by $\langle\delta T\rangle=1.5$ mK at $f_S$ = 2 Hz for a 801-micrometer-thickness silver at $T_0=295$ K. The inset shows the AC
  oscillation amplitude $\langle\delta L\rangle$ with $t_c=25$ s (black curve) and $t_c=25$ s (red curve) for a 14000 s period. (b) Histograms of uncertainty of $\langle\delta L\rangle/L$, $\langle\delta L\rangle_U/L$, in inset of (a). The red line is a
  Gaussian fit with $\langle\delta L\rangle_{U}/L= 6\times 10^{-10}$ standard deviation. (c) Measured
  piezoelectric coefficient $d_{33}$ of 0.5 mm-thick Z-cut LiNbO$_3$ at 305 K.
  We apply an oscillating voltage of different frequencies on the
  sample's thickness-direction, measured its AC piezoelectric length
  change $\delta L_D$, and then calculate $d_{33}$ using
  eq (9). $d_{33,\delta L}$ (black squares) is calculated from
  $\delta L_D$ measured by $\delta L$ and $d_{33, \Delta L}$ (red
  circles) is calculated using the $\delta L_D$ measured from
  $\Delta L$. The crossover between $d_{33, \Delta L}$ and
  $d_{33,\delta L}$ at about 0.1 Hz identifies the feedback
  bandwidth. The black dash line represent the expected $d_{33}$,
  i.e. $d_{33, \delta L}$ measured at sufficiently high frequency well
  above the feedback loop bandwidth. This figure can be used as a
  calibration of our interferometer if the absolute value of length
  oscillation is of interest.}
\end{figure}

For demonstrating the $\langle\delta L\rangle$ resolution of our dilatometer, we apply an AC temperature $\langle\delta L\rangle=1.5$ mK at $f=2$ Hz to a 801-micrometer-thickness silver to induce an extremely small $\delta L$ about $\langle\delta L\rangle=18$pm at $T_0=295$ K. Fig. 3(a) shows the power spectrum density (PSD) of the raw $\delta L$ signal
measured by the interferometer, where the noise spectral density at
frequency $f_S$ is about $3.3\cdot f_S^{-0.5}$ pm/$\sqrt{\mathrm{Hz}}$. The
two peaks corresponds to the first and second harmonic component of
$\delta L$ induced by an $f=$ 2 Hz voltage applied to the AC
heater and the noise spectral density at $f=2$ Hz is about 8 dB re
pm/$\sqrt{\mathrm{Hz}}$, as well as 2.5 pm/$\sqrt{\mathrm{Hz}}$. The inset of Fig. 3(a) shows the measured $\langle\delta L\rangle$ by lock-in technology with $t_c=25$ s in time domain, where $t_c$ is the time constant of our lock-in algorithm, the same with time constant of commercial Lock-In Amplifier. The time constant $t_c$ is a compromise between the measurement time
and the resolution of $\delta L/L$. We can estimate the uncertainty of
$\delta L/L$ is
$\langle\delta L\rangle_U/L=\frac{2.5 \text{pm}/\sqrt{\mathrm{Hz}}}{\sqrt{t_c}\cdot L}$ at $f_s=2$ Hz so $\langle\delta L\rangle_U/L$ is $6\times 10^{-10}$ for the $L=801 \mu$m silver when $t_c=25$ s, as the histogram of data in inset of Fig. 3(a) showed in Fig. 3(b). It is improved by a factor of 2 for every 4-fold increase in $t_c$ or $f_S$ so it can reach remarkable $1\times 10^{-10}$ resolution if we choose $t_c=900$ s. On the other
hand, the change of $T_0$ during the averaging time $t_c$ should be
smaller than the $\langle \delta T\rangle$ so that the temperature resolution of our dilatometer is dependent on $\langle\delta T\rangle$. The
$\langle \delta T\rangle$ can be as low as 5 mK in order to resolve
the fine structures in $\alpha$, e. g. the sharp $\alpha$ peak at the
first order phase transition, see Fig. 4. So in typical measurement , the
sweeping rate of $T_0$, $dT_0/dt$, is usually about 0.1 $\sim$ 1 mK/s,
and $\langle \delta T\rangle$ is about 0.5 K if there is no need for extremely high temperature resolution. Even under the worst of circumstances in which the sample is as thin as $L=200 \mu$m and we should measure at frequency as low as $f=0.25$ Hz for thermal equilibrium of whole sample while AC heating, the resolution of $\alpha$ is still as low as 8$\times 10^{-9}$ K$^{-1}$ when $t_c=$25 s,
and $10^{-10}$ K$^{-1}$ resolution is achievable by increasing
$t_c$ and $\langle \delta T\rangle$. For better measurement condition with thicker sample and higher measurement frequency $f_S$, the $10^{-10}$ K$^{-1}$ resolution of $\alpha$ is easy to achieve.

In addition to its resolution, measuring $\alpha$ as accurate as
possible is also of great interest. The precision in $\alpha$ is
limited by the $\langle \delta L\rangle$ measurement, because the
amplitude of the AC temperature oscillation,
$\langle \delta T\rangle$, is proportional to the AC heating power
$V_{ACH}^2$ and can be measured/calibrated with extremely high
accuracy. We demonstrate the precision of $\langle \delta L\rangle$ by
measuring the piezoelectric coefficient $d_{33}$ of a Z-cut LiNbO$_3$
single crystal with thickness $L=0.5$ mm. The upper and lower surfaces
of sample are coated with 20 nm-thick Au film for gating. An AC
voltage $V_D$ is applied between these two gates to generate a
thickness change $\delta L_D$, which is then measured by the
interferometer as described in previous sections. The $d_{33}$ is then
deduced by:
\begin{equation}\label{eq6}
  d_{33}=\frac{\delta L_D/L}{V_D/L}=\frac{\delta L_D}{V_D}
\end{equation}

Fig. 3(d) shows the measured $d_{33}$ of LiNbO$_3$ at
305K. As described in last section, the sample's thickness change can
either be deduced from the interferometer output, $\delta L$, or from
the feedback voltage, $\Delta L$. $d_{33,\delta L}$ is deduced from
$\delta L$ and $d_{33,\Delta L}$ is obtained from the measurement of
$\Delta L$. At low frequencies $f_S<0.1$ Hz, most of the slowly
varying $\delta L_D$ will be compensated by the feedback voltage, so
that $d_{33,\delta L}$ reduces towards zero. At high frequency
$f_S>0.1$ Hz, the feedback loop cannot response fast enough and
$\delta L_D$ is mostly captured by $\delta L$. At sufficiently high
frequencies $f_S>1$ Hz, $d_{33,\Delta L}$ nearly vanishes so that
$d_{33}=d_{33,\delta L}=5.8$ pm/V, consistent with other
reports\cite{RN214,RN215,RN216,RN217}. In principle, one can extend
the low frequency performance by changing the PID feedback
parameters. However, it is not causing a problem in our $\alpha$
measurement, since the optimized frequency of our in-situ heating
sample holder is about 0.1 $\sim$ 2 Hz.

\section{\label{Thermal expansion measurement of BaFe$_2$As$_2$}Thermal expansion measurement}

Finally, we demonstrate the performance of our dilatometer by studying
the thermal expansion coefficient of BaFe$_2$As$_2$, the parent
compound of the "122" Fe-based superconductors. It is generally
believed to have a first-order antiferromagnetic transition at $T_\mathrm{N}$
around 134K\cite{RN218,RN219}. We choose a sample with thickness
$L=210 \mu$m and weight $m=5$mg. Fig. 4(a) shows the
typically raw data of $\delta T$ and $\delta L$ using $f_S=0.25$Hz at
$T_0\simeq$130K. The phases of these two oscillations are perfectly
aligned so that the $\delta L$ vs. $\delta T$ plot in Fig. 4
(b) exhibit a linear relation whose slope is the $\alpha$. This linear
dependence evidences that the sample and sample holder are always at
thermal equilibrium and have a uniform temperature, because a finite
temperature gradient will result in a delay of the sample temperature
and the measured thermal expansion $\delta L$ from the temperature
variation $\delta T$ taken at the vicinity of the AC heater.

\begin{figure}[!htbp]
\includegraphics[width=0.5\textwidth]{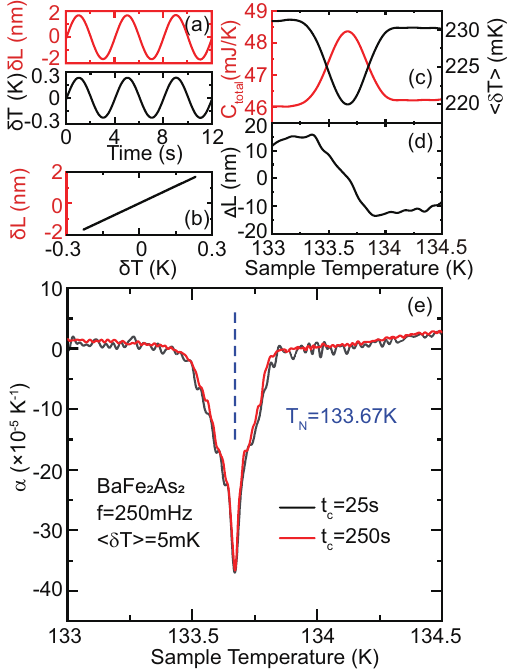}
\caption{\label{Fig4} $\alpha$ measured at $f_S$ = 0.25Hz from a
  BaFe$_2$As$_2$ sample with thickness 210 $\mu$m. (a) The $\delta L$
  and $\delta T$ oscillation at $T_0$ = 130 K. (b) The linear relation
  between $\delta L$ and $\delta T$ of the data in (a). (c) The
  divergence of $\langle \delta T\rangle$ (black) when $T_0$ sweeps
  through the transition temperature. We can calculate the total heat
  capacity $C_{total}$ (red) of heated components. (d) The $\Delta L$
  measured synchronously with (c) data. (e) The AC $\alpha$ measured
  with $f_S$ = 0.25Hz and $\langle \delta T\rangle$ =5 mK using
  averaging time $t_c$ = 25 and 250 s, respectively.}
\end{figure}

In our dilatometer, the sapphire sample holder is nearly thermally
isolated from the rest of the system, so that its temperature change
causes negligible thermal expansion of the structure. The black
curve in Fig. 4(c) represents the measured $\langle \delta T\rangle$
when we sweep the DC temperature $T_0$ through the phase transition
temperature $T_\mathrm{N}$. From the measured
$\langle \delta T\rangle$, we can calculate the total heat capacity
$C_{total}$ of the components that has been isolated together with the
sample holder. According to the known heat capacity of
BaFe$_2$As$_2$\cite{RN220, RN221}, sapphire and copper, we
estimate that the heated structure includes the sapphire plate (15),
sample (14) and the lower plate (16).

In addition, $\langle \delta T\rangle$ exhibits a dip at 133.7 K,
signaling a maximum in $C_{total}$. This is consistent with the
diverging specific heat of BaFe$_2$As$_2$ at its phase
transition \cite{RN220, RN221}. We plot the measured sample's
thickness change $\Delta L$ from the feedback voltage in
Fig. 4(d). The 30-nm $\Delta L$ jump coincides with the
$C_{total}$ peak, as expected from the Ehrenfest and Maxwell
relations.

Thanks to the extremely high resolution, our dilatometer reveals more
features near the phase transition. Fig. 4(e) shows the AC
thermal coefficient $\alpha$ measured using
$\langle \delta T\rangle=5$ mK, as well as sub-10 mK
resolution in temperature. $\alpha$ exhibits a large negative
peak near $T_\mathrm{N}=133.67$ K. The data measured with $t_c=25$ s is plotted in black and the red curve shows the data measured with $t_c=250$ s for better $\langle \delta L\rangle/L$ resolution. The overlapping of the
two traces in Fig. 4(e) demonstrate the stability and accuracy of our
dilatometer. Note that such high resolution in
$\langle \delta L\rangle/L$ and the sub-10-mK resolution in temperature, is achieved simultaneously for a 210-micrometer-thickness sample!

\section{\label{CONCLUSION AND OUTLOOK}CONCLUSION}

In conclusion, we develop a contactless dilatometer based on optical
Michelson interferometer. Our dilatometer is able to measure very thin
and fragile samples with sub-pm-resolution in sample thickness change and
mK-resolution in temperature, simultaneously. We demonstrate its performance by
measuring the antiferromagnetism transition of BaFe$_2$As$_2$. Besides
the thermal expansion coefficient and piezoelectric coefficient
measurement we demonstrate here, our setup can also be used for
magneto-striction, magnetic torque and other measurements where a
length change is introduced by a certain physics process.  It's also
possible to integrate multiple synchronous measurements and in-situ
tuning in the future. Although it's difficult for us now to measure the absolute value of $\alpha$ because of residual AC background thermal expansion of a few nm/K level, both its extremely high resolution, accuracy and
the contactless nature open a new playground for research in condensed
matter physics. This prototype should be upgraded for better performance in the future.

\begin{acknowledgments}
  The work at PKU was supported by the National Key Research and Development Program of China (Grant No. 2021YFA1401900 and 2019YFA0308403), the Innovation Program for Quantum Science and Technology (Grant No. 2021ZD0302602), and the National Natural Science Foundation of China (Grant No. 92065104 and 12074010). The BaFe$_2$As$_2$ sample is supplied by Xingyu Wang,
  Huiqian Luo and Shiliang Li at the Institute of Physics, Chinese
  Academy of Sciences. We thank Mingquan He and Yuan Li for valuable
  discussion.
\end{acknowledgments}

\section*{Data Availability Statement}

The data that support the findings of this study are available from the corresponding author upon reasonable request.

\section*{REFERENCES}
\bibliography{references.bib}

\end{document}